\documentclass[sigconf,screen]{acmart}

\usepackage[show]{chato-notes}
\usepackage{pdflscape}
\usepackage{afterpage}
\usepackage{geometry}
\usepackage{siunitx}
\usepackage{graphicx}
\usepackage{multirow}
\graphicspath{{figs/}}
\usepackage{svg}

\AtBeginDocument{%
  \providecommand\BibTeX{{%
    \normalfont B\kern-0.5em{\scshape i\kern-0.25em b}\kern-0.8em\TeX}}}

\copyrightyear{2021} 
\acmYear{2021} 
\setcopyright{acmlicensed}\acmConference[CHI '21]{CHI Conference on Human Factors in Computing Systems}{May 8--13, 2021}{Yokohama, Japan}
\acmBooktitle{CHI Conference on Human Factors in Computing Systems (CHI '21), May 8--13, 2021, Yokohama, Japan}
\acmPrice{15.00}
\acmDOI{10.1145/3411764.3445082}
\acmISBN{978-1-4503-8096-6/21/05}

\begin{document}

\title{Clandestino or Rifugiato?\\Anti-immigration Facebook Ad Targeting in Italy}
\titlenote{In Italian, ``clandestino'' means ``clandestine'' and ``rifugiato'' means ``refugee''.}

\author{Arthur Capozzi}\authornote{After the first author, the author names are in alphabetical order.}
\affiliation{\institution{ISI Foundation}\city{Turin}\country{Italy}}
\email{arthur.capozzi@isi.it}

\author{Gianmarco De~Francisci~Morales}
\affiliation{\institution{ISI Foundation}\city{Turin}\country{Italy}}
\email{gdfm@acm.org}

\author{Yelena Mejova}
\affiliation{\institution{ISI Foundation}\city{Turin}\country{Italy}}
\email{yelena.mejova@isi.it}

\author{Corrado Monti}
\affiliation{\institution{ISI Foundation}\city{Turin}\country{Italy}}
\email{corrado.monti@isi.it}

\author{Andr\'{e} Panisson}
\affiliation{\institution{ISI Foundation}\city{Turin}\country{Italy}}
\email{andre.panisson@isi.it}

\author{Daniela Paolotti}
\affiliation{\institution{ISI Foundation}\city{Turin}\country{Italy}}
\email{daniela.paolotti@isi.it}

\renewcommand{\shortauthors}{Capozzi et al.}

\begin{abstract}

Monitoring advertising around controversial issues is an important step in ensuring accountability and transparency of political processes.
To that end, we use the Facebook Ads Library to collect 2312 migration-related advertising campaigns in Italy over one year.
Our pro- and anti-immigration classifier (F1=0.85) reveals a partisan divide among the major Italian political parties, with anti-immigration ads accounting for nearly 15M impressions.
Although composing 47.6\% of all migration-related ads, anti-immigration ones receive 65.2\% of impressions.
We estimate that about two thirds of all captured campaigns use some kind of demographic targeting by location, gender, or age.
We find sharp divides by age and gender: for instance, anti-immigration ads from major parties are 17\% more likely to be seen by a male user than a female.
Unlike pro-migration parties, we find that anti-immigration ones reach a similar demographic to their own voters.
However their audience change with topic: an ad from anti-immigration parties is 24\% more likely to be seen by a male user when the ad speaks about migration, than if it does not.
Furthermore, the viewership of such campaigns tends to follow the volume of mainstream news around immigration, supporting the theory that political advertisers try to ``ride the wave'' of current news.
We conclude with policy implications for political communication: since the Facebook Ads Library does not allow to distinguish between advertisers intentions and algorithmic targeting, we argue that more details should be shared by platforms regarding the targeting configuration of socio-political campaigns.

\end{abstract}


\maketitle

\section*{Interactive visualization}
\href{http://migration-ads-observatory.isi.it/}{migration-ads-observatory.isi.it}

\section{Introduction}
According to the International Organization for Migration, almost 130k migrants have arrived in Europe in 2019.\footnote{\url{https://migration.iom.int/europe?type=arrivals}}
Global migration is a systemic challenge for Europe~\cite{joppke1998challenge}, and for Italy in particular which is at the forefront of the Mediterranean route~\cite{frontex2020migratory}.
Migration is thus, unsurprisingly, a central issue in European and Italian politics~\cite{human2020events}.

At the same time, Europe and the world in general have seen a resurgence of nationalism with a populist derive, such as what seen in USA, Brazil, Philippines, Turkey, UK, Hungary, and Italy to name a few~\citep{smith2013nations}.
Nationalist parties often promote nativist positions, which highlight the negative effects of migration, and emphasize the loss of cultural identity~\citep{falk2010invasion}.
Indeed, migration is a `hot-button' issue that often sparks controversies in the political conversation across the aisle.
In this work, we focus on the Italian debate around migration, which has been of paramount importance in recent history.\footnote{\url{https://www.bbc.com/news/world-europe-43167699}\\ \url{https://www.politico.eu/article/italy-immigration-debate-facts-dont-matter}}

Some of the success of populist parties has been attributed to their embrace of the new communication technologies available on the Web and social media~\citep{beer2019landscape}.
In Italy, the Five Star Movement (\emph{Movimento 5 Stelle}, M5S)~\citep{natale2014web} is a famous example of this phenomenon, with its focus on anti-establishment rhetoric paired with an organizational focus around digital communication technologies.
For instance, they favor direct ``e-democracy'', and advocate putting ``citizen in power'' by using the Internet~\cite{grillo2013citizen}.
Also Matteo Salvini, leader of Lega and described by~\citet{bulli2018immigration} as ``one of the first political entrepreneurs of anti-immigration sentiments in the Italian arena'', has been effective in using social media to spread his message.%
\footnote{See for instance \href{https://www.theatlantic.com/international/archive/2019/09/matteo-salvini-italy-populist-playbook/597298/}{The Atlantic, \emph{``The New Populist Playbook''}}, %
\href{https://www.reuters.com/article/us-italy-politics-salvini-socialmedia/chestnuts-swagger-and-good-grammar-how-italys-captain-builds-his-brand-idUSKCN1MS1S6}{Reuters.com, \emph{``Chestnuts, swagger and good grammar: how Italy's 'Captain' builds his brand''}} and %
\href{https://www.theguardian.com/news/2018/aug/09/how-matteo-salvini-pulled-italy-to-the-far-right}{The Guardian, \emph{``How Matteo Salvini pulled Italy to the far right''}}. %
}
In particular, he is active on Facebook, where he is the most popular politician in Europe by number of followers.\footnote{\href{https://www.lastampa.it/politica/2020/01/13/news/salvini-e-il-record-della-propaganda-social-su-facebook-4-milioni-di-seguaci-1.38320663}{La Stampa, \emph{``Salvini e il record della propaganda social''}}}

The advent of social media, and the user profiling it brings, has enabled in fact direct and personalized communication, in sharp contrast with traditional media (TV and newspapers) which focus on mass communication.
This \emph{micro-targeting} feature is highly controversial and has caused a wide backlash, especially when it comes to political advertising (e.g., Facebook~\cite{hern2019facebook}).

Given this socio-political context, our main research goal is to study the \emph{political messaging around migration in Italy via advertising campaigns run on Facebook}, the largest social media platform in the world.
In particular, we analyze the stances toward migration of the major parties on the Italian stage, and look specifically for evidence of micro-targeting in the way the campaigns reach demographics of certain gender, age, and location.
We find that different parties have different demographic foci, and the target audience peaks during events like elections.
By developing a supervised classifier, we categorize over 2 thousand campaigns as to the stance on the migration issue: pro or anti. 
These stances, we find, align neatly along the party lines, each having distinct audience in terms of demographic groups reached.
For example, nationalist anti-migration parties focus on more male audience for their migration ads compared to their normal targets, and compared to the other parties.
Further, we contextualize the temporal dynamics of the migration ad campaigns in the mainstream news media covering ongoing events such as the elections, changes in the government, and migrant boat arrivals.
We find that the ads ``ride the wave''~\cite{ansolabehere1994riding} of the media attention to ongoing events, and especially so for the anti-immigration ones.
Finally, we discuss the advantages and limitations of this methodology, future integration of online data in political discourse analysis, and design implications for platforms to increase transparency and accountability.

\section{Related Work}

Migration is one of the most polarizing issues in European politics, at the core of the platforms of many right-wing populist parties.
In the case of Italy, this sphere is represented by Lega party~\cite{ivaldi2017varieties}, which obtained 34.3\% of the votes in the 2019 European elections, and Brothers of Italy, which obtained 6.4\%.
Lega, under the leadership of Matteo Salvini, is in fact keeping a strong anti-immigration and nativist focus, adopting ``stop the invasion'' as a slogan, calling for immediate  repatriations, and depicting Islam as a threat to Italian Christian identity~\cite{ivaldi2017varieties}.
As the interior minister in June 2018, Salvini declared Italian ports closed to NGO ships rescuing migrants~\cite{cusumano2020deep}.
The other two major parties in Italy at the 2019 European elections were the Democratic Party (PD), which obtained 22.7\% of the votes, and the Five Star Movement (M5S), with 17.1\%.
The former, in government from 2013 to 2018, is often the target of right-wing attacks on immigration; PD adopted a strategy of ``setting up a decentralized system for the management of asylum requests''~\cite{diamond2019italian} and reducing migrant flows to Italy~\cite{guardian2017}.
M5S, described as anti-establishment and populist, has been hard to describe on the left-right spectrum~\cite{mosca2019beyond}.
\citet{emanuele2020times} found that ``M5S voters are leftist on the economy but quite close to right-wing voters on Europe and immigration''.
This is reflected by some of the M5S press releases; for instance, they accused NGOs of increasing illegal immigration by rescuing migrants at sea~\cite{coticchia2020populist}.
These two parties, PD and M5S, formed a new government in August 2019, after Lega breached their governing coalition with Five Star Movement, hoping to trigger new elections~\cite{telegraph2019}.
In this study, we attempt to capture the publicity campaigns that have accompanied this highly diverse and dynamic political situation. 

In the field of Human-Computer Interaction (HCI), advertising has been studied through the lens of persuasive technologies \cite{fogg2002persuasive}.
The user's perception of information shown on the Internet, and the resulting reception to the persuasive message, have been linked to the user's moral values and worldview~\cite{magee2010perceived}, physical state at the time of interaction~\cite{lee2011mining}, and a myriad of cognitive biases~\cite{caraban201923}.
Simultaneously, intense research in political communication has shown the importance of interaction with diverse media during the deliberation process \cite{semaan2014social}, yet has also revealed hyper-personalization that results in ideological ``echo chambers'' \cite{garimella2018political}.
Systemically, the social media platforms play an increasingly important role in the political messaging ecosystem, giving rise to what \citet{woolley2018computational} call ``computational propaganda'' -- a combination of manual curation and algorithmic optimization in order to maximize the propagation of a message.
As online advertising becomes increasingly sophisticated, political scientists are worried that the responsibility for safeguarding of democratic process is being ceded to commercial actors ``who may have differing understandings of fundamental democratic norms''~\cite{dommett2019political}.
For example, \citet{speicher2018potential} show that Facebook Advertising platform provides features such as creation of custom audiences via personally identifiable information (PII), narrowing of the audience via potentially sensitive attributes (such as ethnic affinity), and ``look-alike'' audience matching that all can be used for discriminatory advertising.
Unlike traditional media, advertising is increasingly used to boost content and engage the viewer, with the aim to achieve ``virality'' and further increase campaign's audience (for instance, Donald Trump's 2016 campaign achieved favorable virality on Twitter, nearly twice as much as Clinton's~\cite{darwish2017trump}).
Increasingly, social media platforms, and especially Facebook, has been under fire for providing a platform for divisive messaging that is seen as harmful to democracy~\cite{vaidhyanathan2017facebook,entous2017russian}, in response to which the platforms tightened their policies, such as not running any political ads during the week before the 2020 U.S. Presidential election~\cite{facebook2020steps}.
Another response to such criticism was the release of Facebook Ads Library, which makes available to public scrutiny the political advertisements and their audiences.
Here, we use machine learning to automatically extract the political leaning in thousands of ads. Combining it with the audience information, we attempt to gauge to what extent each side engages in audience targeting, and discuss the adequacy of the data Facebook Ads Library exposes. 

Further, we attempt to contextualize the migration-related advertising with respect to mainstream news media coverage. 
The relationship between political advertising and news has been investigated under the lens of communication theory, and in particular within the context of agenda-setting theory.
If media is able to influence which topics the public will consider important~\cite{mccombs2002news}, is political advertising on media also able to affect this agenda?
This question was tackled among the firsts by~\citet{roberts1994agenda}: they measure the correlation over time between the issues suggested by TV political advertising from the 1990 Texas gubernatorial election, and the distribution of topics in the news.
They find a strong correlation between ads and the following news agenda, both in newspapers and television, while the opposite relation (news influencing ads) is essentially zero.
The authors interpret this observation as a direct response from news organizations to ``the overall campaign agenda, for which the advertising is a significant and clear-cut manifestation'', in the same way as they cover candidates speeches and press releases.
\citet{ansolabehere1994riding} introduced two hypothesis to better understand the effectiveness of such advertising: the \emph{issue-ownership} hypothesis, where candidates gain the most from advertising on issues over which they can claim "ownership", and the \emph{riding-the-wave} hypothesis, where the candidates' campaigns are more effective when mentioning the same topics as the current news coverage.
More recently, \citet{kluver2016setting} categorized press releases from German parties and measured the correlation of their topics with issues identified as important by voters through surveys: they find that political parties tend to respond to voters, and not the other way around.
This finding supports the riding-the-wave hypothesis: political parties ``emphasise policy issues that are salient in the minds of citizens''.
A similar result is found by~\citet{thorson2018attention} during the Clinton-Trump campaign, by measuring the attention of voters to political advertising through surveys.
They find that when a salient political event is heavily discussed in newspaper, there is a measurable increase in attention to ads.
In our work, we adapt these questions to the social network era, and discover that Facebook political advertising on the topic of migration is influenced by how much the news is discussing the same topic.

\section{Data}

\begin{table}[t] %
\footnotesize
\begin{center}
  \caption{Occurrences for each considered theme in the \textsc{GDELT} News data set, with and without counting the repetitions inside the same news (one theme can occurs more than one time in a news). Total shows the number of migration-related themes and the number of news articles included.}
\begin{tabular}{lrr}
\toprule
\textsc{GDELT} Theme & \hspace{-3em} Occurrences & News articles \\
\midrule
EPU Cats Migration Fear Fear                       &           385722 &             285502 \\ 
EPU Cats Migration Fear Migration                  &           111271 &              73105 \\ 
Wb 2836 Migration Policies And Jobs                &           102322 &              71356 \\ 
Immigration                                        &            69210 &              47451 \\ 
Wb 2837 Immigration                                &            47584 &              32266 \\ 
Tax Fncact Immigrants                              &            25780 &              17503 \\ 
Wb 2844 Emigration                                 &             7008 &               4765 \\ 
Tax Fncact Immigrant                               &             5362 &               3999 \\ 
Discrimination Immigration Xenophobic              &             2507 &               2082 \\ 
Discrimination Immigration Xenophobia              &             1547 &               1234 \\ 
Discrimination Immigration Antiimmigration         &              354 &                327 \\ 
Discrimination Immigration Antiimmigrant           &              314 &                283 \\ 
Discrimination Immigration Ultranationalist        &              256 &                226 \\ 
Human Rights Abuses Forced Migration               &               92 &                 89 \\ 
Soc Massmigration                                  &               85 &                 82 \\ 
Discrimination Immigration Against Immigrants      &               39 &                 36 \\ 
Wb 2204 In Migration                               &               25 &                 24 \\ 
Tax Fncact Migrant Worker                          &               19 &                 15 \\ 
Wb 2729 Migrant Workers                            &               19 &                 15 \\ 
Discrimination Immigration Anti Immigration        &                3 &                  3 \\ 
Discrimination Immigration Ultra Nationalist       &                1 &                  1 \\ 
Discrimination Immigration Attacks Against Im-nts &                1 &                  1 \\ 
Wb 1602 Returning Migrants                         &                1 &                  1 \\ 
\bottomrule
Total & 759523 & 385945 \\
\bottomrule
\end{tabular}
\label{table:theme-occurrences}
\end{center}
\end{table}

\subsection{Facebook Ads Library}

In this study, we use the Facebook Ads Library,\footnote{\url{https://www.facebook.com/ads/library}} a tool made available by Facebook in March 2019 to surface past and ongoing advertising campaigns about ``social issues, elections, or politics''.
Using the tool's API,\footnote{\url{https://www.facebook.com/ads/library/api}} we query the platform on March~30, 2020 to collect advertisements originating from Italy, containing keywords (in Italian) related to immigration and refugees. 
We find only two ads in March 2019, with the bulk of the ads starting in April, which suggests that the tool was not properly registering ads yet at the beginning of the time period.
Information about each of the \num{2312} ads we collect includes an ID, title, the text of the ad, a URL, as well as when it was created, the span of time it ran, and the cost of the ad (given as a range).
We show five examples of the ads we find in Table~\ref{table:example_labels}, together with the labels we will define later in Section~\ref{sec:annotation-guidelines}.
Each ad is associated with a Facebook page, which in turn can be associated with an individual, company, or organization.
We identify \num{733} unique pages in this dataset.
Crucially, the platform also provides information on the users who saw the ad (so-called ``impressions''): the total number of impressions (given as a range), as well as broken down by gender (male, female, and unknown),  7 age groups, and location down to region.
For those values that come in a range (cost and impressions) we take the average of the end points of the range, and for open-ended ranges we take the known closed end point.
For a further description of the dataset see~\cite{FacebookAds2020}. %

As the focus of this study is the political discourse around migration, we examine the pages to find those in the political arena.
To do this, we break down the title of pages in all possible n-grams and use them to query WikiData,\footnote{\url{https://www.wikidata.org}} which, among other categories, returns entities that are parties, politicians, journalists, and NGOs.
After manually disambiguating duplicate matches, we also match the pages to a list of local Italian politicians.\footnote{\url{https://dait.interno.gov.it/elezioni/open-data/dati-amministratori-locali-carica-al-31-dicembre-2018}}
This way, we are able to identify 249 pages as politicians and 53 as parties (including regional branches).
Among these pages, we find those affiliated with one of the five major parties in Italy: Democratic Party (Partito Democratico, PD), League (Lega), 5 Stars Movement (Movimento 5 Stelle, M5S), Brothers of Italy (Fratelli d'Italia, FdI), and Italy Alive (Italia Viva, IV) -- which altogether cover 208 pages.
To better understand overall political advertising, for each page we query the platform without any keyword constraint, resulting in \num{17014} ads that we use as a general political advertising baseline.

\subsection{\textsc{GDELT} News}

We also compare the ads above to the simultaneous Italian news coverage from the Global Database of Events, Language and Tone (\textsc{GDELT})\footnote{\url{http://www.gdeltproject.org}}~\cite{leetaru2013gdelt}. %
\textsc{GDELT} data are collected from real-time translations of worldwide news into 65 languages and updated every 15 minutes.
This way, we collect URLs published in Italian language along the same time period as the Facebook ads.
The \textsc{GDELT} platform first translates each article from Italian to English, then extracts a list of themes from the translated text, along with other text entities such as counts, people names, organization names, locations, emotions, relevant imagery, video, and embedded social media posts.
We keep only the article's date, URL (\texttt{DocumentIdentifi- er}), and the list of themes (\texttt{V2THEMES}).

We then proceed to identify, among all themes present in the selected data, those that are related to the migration debate.
For reproducibility, Table~\ref{table:theme-occurrences} reports the list of themes, with their number of occurrences in the data.
Note that the most common themes extracted by \textsc{GDELT} are related to Economic Policy Uncertainty (EPU) categories, also used to build the economic policy uncertainty index by~\citet{baker2016measuring}.

\section{Methods}

\begin{table*}[htb!]
\footnotesize
\begin{center}
\begin{tabular}{ p{5cm} p{3cm} p{5cm} p{3cm}  }
\toprule

\raisebox{-\totalheight}{\includegraphics[width=4.9cm]{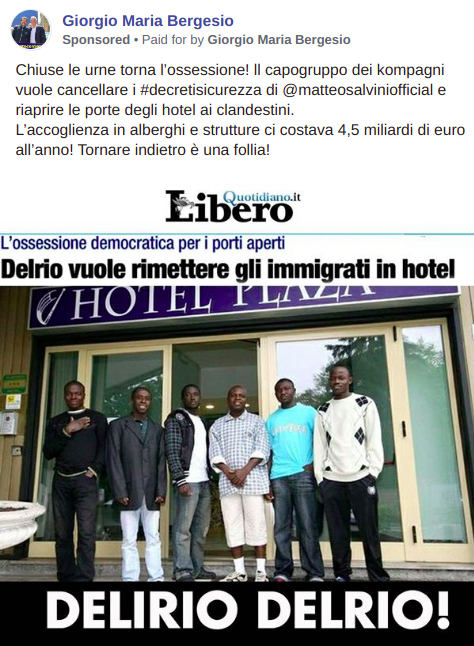}} &
\textbf{Strongly agree: }
\emph{After closing the polls, the obsession returns! This pinko group leader wants to abolish Salvini's decrees and reopen the doors of hotels to illegal immigrants. Hospitality in hotels and other structures cost us 4.5 billion euros per year! Going back is madness!} &

\rule{0pt}{4ex}
\raisebox{-\totalheight}{\includegraphics[width=4.9cm]{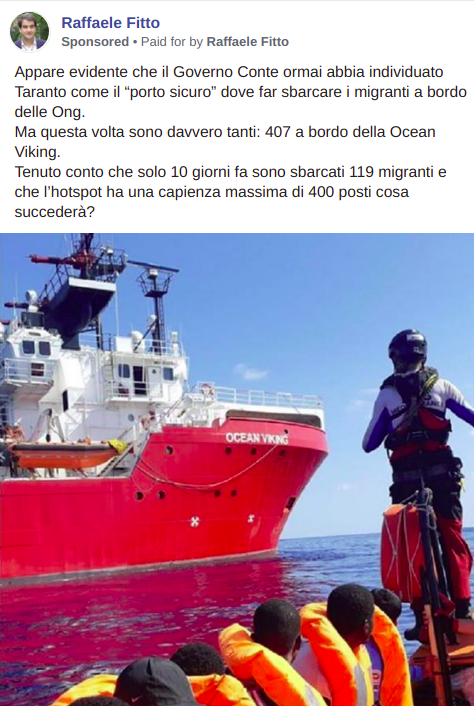}} &
\textbf{Slightly agree: }
\emph{It is clear that the government led by Conte has now identified Taranto as the "safe harbor" where migrants aboard NGOs can be landed. But this time they are truly many: 407 aboard the Ocean Viking.
Taking into account that only 10 days ago 119 migrants landed and that the hotspot has a maximum capacity of 400 places, what will happen?} \\

\rule{0pt}{4ex}
\raisebox{-\totalheight}{\includegraphics[width=4.9cm]{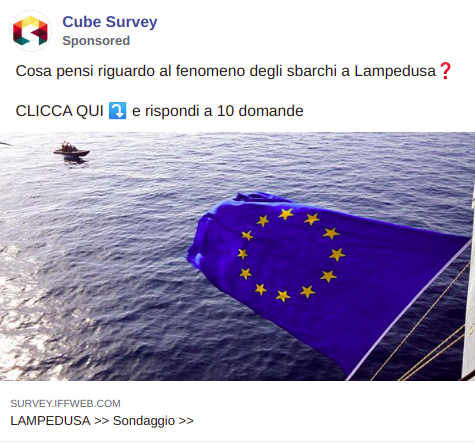}} &
\textbf{Neutral: }
\emph{What do you think about the phenomenon of boat arrivals in Lampedusa? Clink here and answer 10 questions.} 
& 
\multirow{2}{*}{
\raisebox{-\totalheight}{\includegraphics[width=4.9cm]{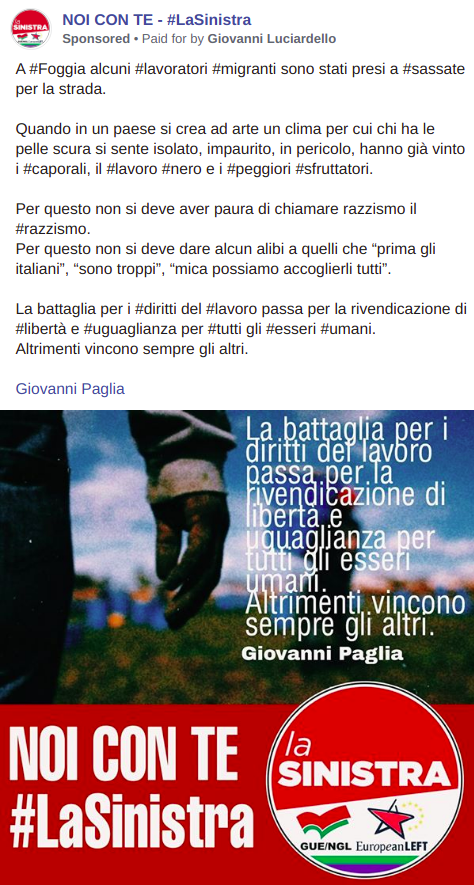}}} &
\multirow{2}{3cm}{ \textbf{Strongly disagree: } \emph{In Foggia some migrant workers were stoned on the street. When they create in the country a climate where those with dark skin feel isolated, afraid and in danger, then gangmasters, illegal work and the worst exploiters have already won. For this reason one should not be afraid to call racism racism. For this reason, no alibi should be given to those saying ``Italians first'', ``they are too many'', ``we can't welcome them all''. The battle for the rights of labour goes hand in hand with the claim of freedom and equality for all human beings. Otherwise the others always win.} } \\

\raisebox{-\totalheight}{\includegraphics[width=4.9cm]{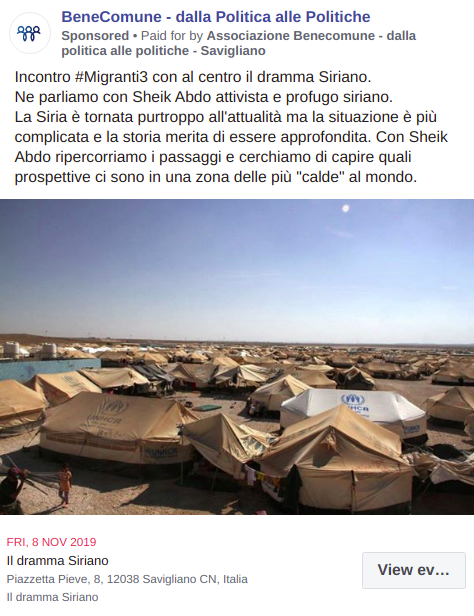}} \vspace{0.5cm}&
\textbf{Slightly disagree: }
\emph{Meeting} \#Migranti3\emph{, focusing on the Syrian drama. We will discuss it with Sheik Abdo, activist and Syrian refugee. Unfortunately, Syria has returned in the news, but the situation is more complicated and this story deserves to be told. With Sheik Abdo, we retrace its steps and try to understand what prospects can we see in one of the most dangerous areas in the world.} & \\

\bottomrule
\end{tabular}
\caption{Examples of ads spanning the spectrum of stances on migration, with English-language translation of ad text, together with the labels defined in Section~\ref{sec:annotation-guidelines}.}
\label{table:example_labels}
\end{center}
\end{table*}

Our first goal is to analyze the attitudes in favor or in opposition to migration, as expressed in the ads that Italian Facebook users may see.
In order to do so, we build a supervised ad classifier based on an annotated dataset.
To develop this approach, we design a labelling task to build a data set of ads with known stances.
At the same time, as the data is retrieved via keywords, we ensure the content is relevant to the topic of migration by labelling irrelevant ads, and building a separate supervised classifier specifically for this task.

\subsection{Annotation guidelines}
\label{sec:annotation-guidelines}

In order to fit the ads attitudes into a well-defined schema, we ask annotators, recruited through convenience sampling, whether the message of the ad agrees with the following main statement: \emph{our country has been made a worse place to live by people coming to live here from other countries.}
To measure such agreement, we use a 5-step Likert scale, going from ``strongly disagree'' to ``strongly agree''. 
This main statement -- the one we measure agreement with -- follows those typically used in public opinion surveys about immigration.
For instance, it has been used by European Social Surveys to measure attitudes towards migration~\cite{heath2014attitudes}.
They include this statement in their core ones as it ``provides an overall measure of support for, or opposition to, immigration''.
As such it has been used by other analysis on opinions about immigration~\cite{sides2007european}.

As in those studies, we employ a Likert scale in order to distinguish different degrees of support.
This strategy has been also recommended by scholars in similar annotation processes for machine learning:~\citet{poletto2019annotating} encourage to use a rating scale to annotate hate speech in text, in order ``to avoid a binary choice that could leave too many doubtful cases'', and produce a higher quality data set.

A key part of the guidelines we designed for annotators is how to treat \emph{implicit} agreement with the given statement.
The importance of implicit statements in text analysis is well-known~\cite{van2015good}.
\citet{habernal2017argumentation} found that 48\% of claims in their corpus of user-generated web argumentative discourses are implicit.
Therefore, while we categorize ads that explicitly state that Italy has been made worse by immigrants as ``strongly agreeing'', we instructed annotators to account for ads that \emph{implicitly} suggest that vision to be labelled as ``slightly agreeing''.
We employ the same coding rule also in reverse: ads that implicitly suggest that Italy has \emph{not} been made worse by immigrants should be labelled as ``slightly disagree''.
In order to clearly communicate this coding protocol to annotators, we also attached example sentences for each label.
We can summarize our guidelines for each label as follows:

\begin{itemize}
    \item \emph{Strongly agree} (coded as 5 in the following). The ad explicitly claims Italy has been made worse by immigrants. For example, ads expressing the following views strongly agree with the given statement:

    \begin{itemize}
        \item ``Immigrants bring degradation; they threaten public security; they represent a cost for our economy.''
        \item ``Immigrants take public resources away from non-immig- rant population, making it harder to help the non-immig- rant population.''
        \item ``We need to avoid at any cost that immigrants arrive on national territory.''
        \item ``Immigration causes shady business activities, especially from NGOs.''
        \item ``We need to fight laws that might encourage immigration.''
    \end{itemize}

   \item \emph{Slightly agree} (4). The ad \emph{implicitly} or \emph{marginally} suggests that Italy has been made worse by immigrants. %
    Views implying one of the above views, but without clearly stating them, should be labelled as slightly agree.

    \item \emph{Neutral} (3). The ad is not stating, suggesting or implying anything about whether Italy has or has not been made worse by immigrants. 
    
    \item \emph{Slightly disagree} (2). The ad implicitly or marginally suggests that Italy has \emph{not} been made worse by immigrants.
    
    \item \emph{Strongly disagree} (1). The ad clearly claims Italy has not been made worst by immigrants.
    Examples include: ``we need laws that make immigrants life easier'', ``inclusion of different national groups improves society'', ``we need to fight discrimination and hate against immigrants''.
    \item \emph{Not relevant}. As mentioned, we take the chance of this labelling to refine the keyword-based approach we employed to select relevant ads: annotators had therefore the ``irrelavant'' label as a possible choice.
\end{itemize}

Table~\ref{table:example_labels} shows five examples of ads that have been unanimously evaluated by annotators, one for each label.
The first example strongly agrees with the main statement, as it claims that immigrants represent a huge cost for taxpayers. 
The second one seems to implicitly agree with the main statement: it claims that incoming migrants are too many and rhetorically asks what will happen, hinting at possible harm.
The last two ads instead implicitly and explicitly (respectively) agree with the opposite statement--i.e., migrants are not making the country worse.
The ``slightly disagree'' one advertises a meeting with a migrant to understand the drama they are escaping from; the ``strongly disagree'' ad explicitly condemns hate against migrants and denounces their work exploitation.

After designing these guidelines, we recruited a set of twelve people proficient with the language and familiar with the Italian political context to label a sample of 200 ads from our data set.
We assigned to each annotator a set of 50 ads, so that each ad receives three distinct labels.
We made the text of each ad available to each annotator directly on the screen, while other information (such as author, image, or date of the ad) was available via a provided link.

\begin{figure}[t]
    \centering
    \includegraphics[width=0.65\linewidth]{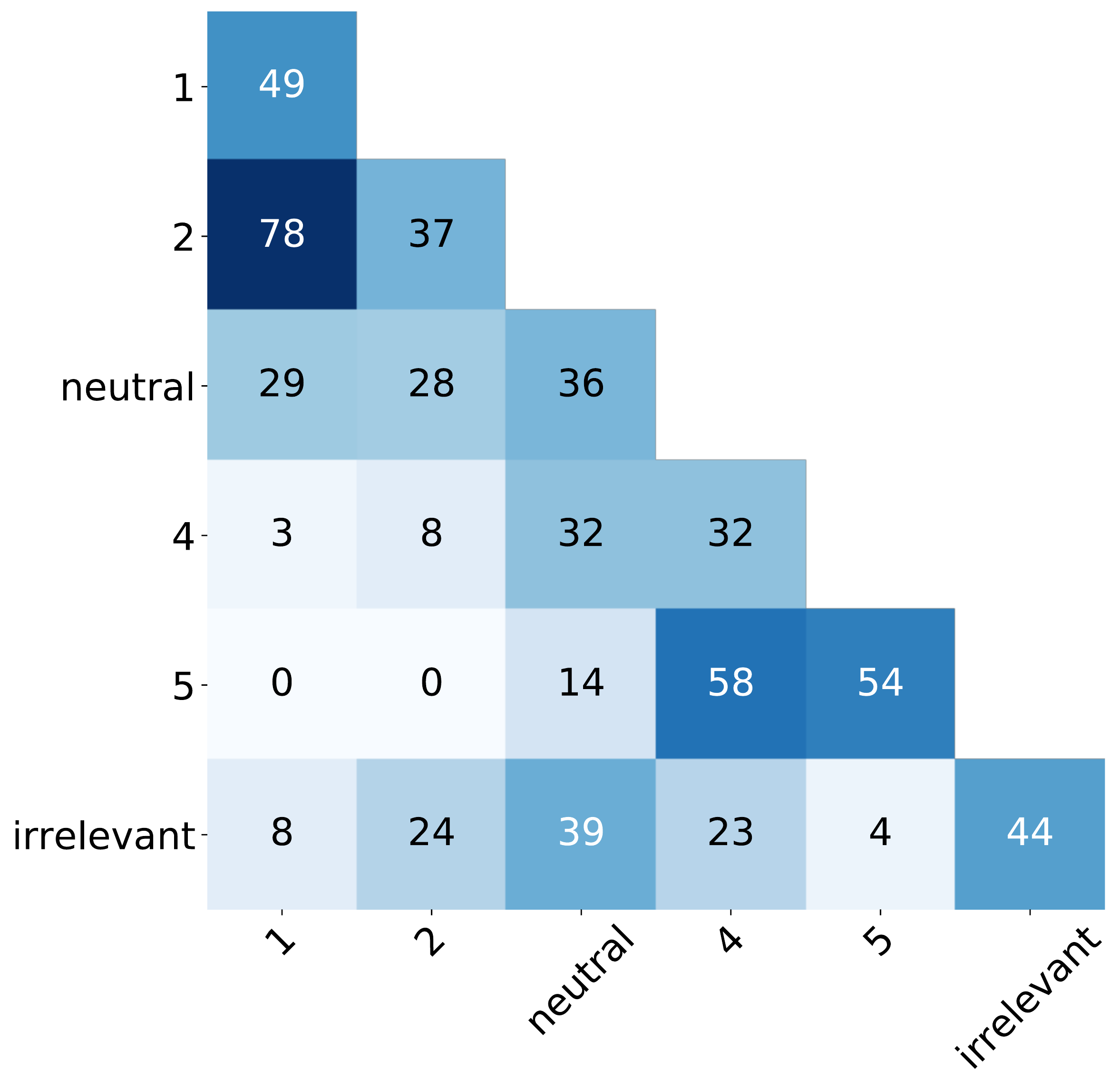}
    \vspace{-0.3cm}
    \caption{Confusion matrix for inter-annotator agreement among 12 annotators on 200 ads (3 evaluation per ad), with the 1-to-5 labels described in Section~\ref{sec:annotation-guidelines}, plus the irrelevant label.}
    \label{fig:confusion-matrix-labels}  
    \Description{Annotators agree for the most part; the more confused label pairs are 1 with 2 and 4 with 5.}
    \vspace{-\baselineskip}
\end{figure}

Since we recognize that our guidelines for labelling involve a degree of arbitrariness in corner cases, we measure the inter-annotator agreement to quantify their objectiveness. 
We remove the 26 ads (13\%) that were judged as non-relevant by at least two annotators, and compute 
Krippendorf's $\alpha$~\cite{krippendorff2011computing} between the remaining five labels.
Krippendorf's $\alpha$ is a measure of agreement that is zero when labels are seemingly random, negative if there is systemic disagreement, and equal to one if the labels agree perfectly.
We use this measure as it can take into account ordinal labels--i.e., assigning maximum distance between labels 1 and 5, and lower distance for closer labels. 
For our five labels, we obtain $\alpha=0.76$ ($n=174$), which indicates a substantial agreement despite the subjectivity of the task.
Henceforth, we restrict our interest to the two opposite polarities (grouping $\{1, 2\}$ and $\{4, 5\}$). For these two groups we have 150 ads, and we obtain $\alpha=0.92$, showing that the two ideological sides can easily be distinguished in this data by annotators.
In Figure~\ref{fig:confusion-matrix-labels}, we show the confusion matrix obtained by counting all the possible pair of labels obtained by each ad.
In the next section, we show how we use these labels to build a classifier, in order to generalize them to all the data set.

\begin{figure}[t]
    \centering
    \includegraphics[width=0.65\linewidth]{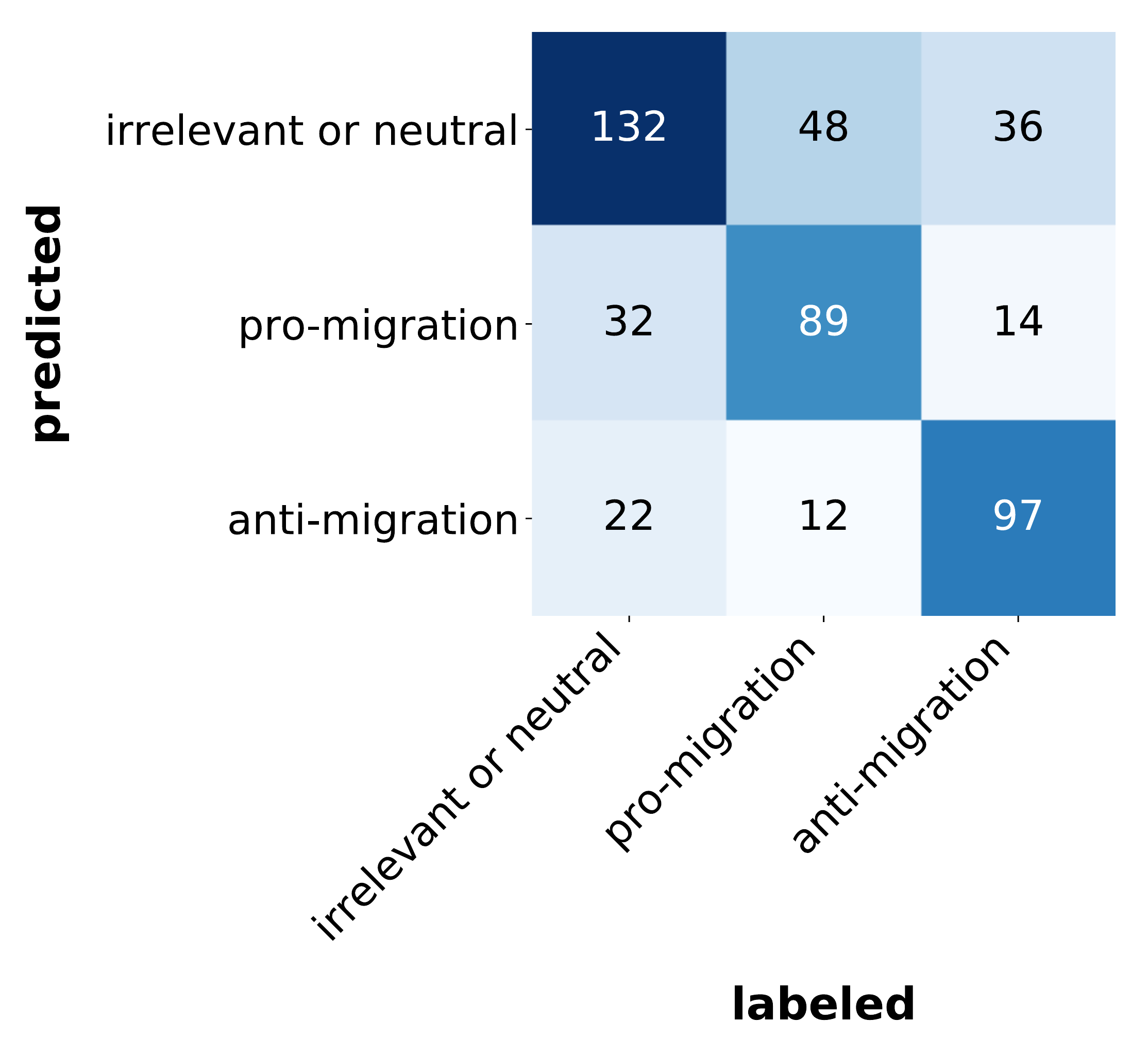}
    \caption{Confusion matrix for the two-stage 3-class classifier, evaluated using 10-fold cross-validation on a final set of 482 ads.}
    \Description{Confusion matrix between the three labels irrelavant/neutral, pro-migrant and anti-migrant, showing that our pipeline is accurate (F1=0.85).}
    \label{fig:confusion_matrix_classifier}
    \vspace{-\baselineskip}
\end{figure}

\subsection{Classifier design}

Using the annotated data described above, we train two classifiers: first to select ads with a stance on migration (``relevance'' classifier), and second to distinguish between the two ideological stances (``leaning'' classifier).
For the relevance classifier, we group ads into two classes: relevant ads with a clear stance from labels $\{1, 2, 4, 5\}$, and ads from label 3 (neutral) and irrelevant ads which express no stance. We consider neutral ads as irrelevant because we are interested in ads that promote a political agenda or view on migration. Moreover, we show in Figure~\ref{fig:confusion-matrix-labels} that annotators are often confused between the two labels, suggesting that the two concepts are partially overlapping.
For the leaning classifier, we group $\{1, 2\}$ -- \emph{pro-migration} henceforth, versus $\{4, 5\}$ -- \emph{anti-migration}.
Since for the relevance task the two classes are unbalanced (141 relevant vs 49 irrelevant), we extend the data for this task by adding 92 ads that are deemed irrelevant according to our keyword-based approach.
In this way, we obtain 141 relevant ads and the same number for irrelevant ones.

We employ 10-fold cross-validation to test the performance of five classifiers: Logistic Regression, Multinomial Naive Bayes, Random Forest, Support Vector Machines with linear and RBF kernels, each optimized via grid search of the appropriate parameter space.
Guided by the F1 measure (harmonic mean of precision and recall), we choose the best performing classifiers for each task: the Random Forest for relevance classification with $F1=0.74$, and Multinomial Naive Bayes for leaning with $F1=0.85$.
The best performing parameters for Random Forest correspond to using 100 estimators, while the minimum number of samples required to split a node is set to 3.
For Multinomial Naive Bayes, we obtain a smoothing parameter $\alpha=0.4$. 

To further assess the performance of the selected classifiers, we obtain additional annotations for 200 ads from 4 of the authors.
Since we obtained a strong agreement between annotators in our first data set, for this extended data annotation we assign a single annotation per ad.
Testing the above classifiers on this new data, we obtain $F1=0.76$ for relevance and $0.85$ for leaning classification.
These results suggest that there is no overfitting on the training data.
Figure~\ref{fig:confusion_matrix_classifier} shows the confusion matrix of the pipeline obtained from these two binary classifiers, over a 10-fold cross-validation on all the labeled data.
While there is some overlap between irrelevant/neutral ads and the other labels, the confusion between pro- and anti-migration stances is minimal.

In the following, we use classifiers trained on all the labeled data.
Further, in support of reproducibility, we release our annotated data set to the public.\footnote{\url{https://github.com/PotenteOpossum/Clandestino-or-Rifugiato-Anti-immigration-Facebook-Ad-Targeting-in-Italy}}

\section{Results}

By applying our leaning classifier on the relevant ads we collected, we can perform different analyses on pro- and anti-immigration messages in Facebook ads.
Specifically, we focus on the following research questions:

\begin{enumerate}
  \item Which messages are most distinctive of pro- and anti-immig- ration attitudes in ads? (Section~\ref{sec:descriptive})
  \item Which demographics are reached by each advertiser and their message?  (Section~\ref{sec:audience-demographics})
  \item Is this targeting intentional? (Section~\ref{sec:ads-targeting})
  \item Do ads follow mainstream news, or on the contrary, set agenda for the news? (Section~\ref{sec:news})
\end{enumerate}

\subsection{Descriptive insights}
\label{sec:descriptive}

First, we inspect the classifier we built in order to assess whether the features used for prediction follow our intuition, and if they can give us insights about the lexicon used by each side.

\begin{table}
  \centering
  \caption{SVM coefficients for the most important features.
  Left: top 10 features with smallest; Right: top 10 with largest.}
  \label{tab:ads_example_3}
  \footnotesize
  \begin{tabular}{@{}llrllr@{}}
    \toprule
    \multicolumn{3}{c}{\textbf{Anti-migration}} & %
    \multicolumn{3}{c}{\textbf{Pro-migration}} \\
    \cmidrule(llr){1-3} \cmidrule(llr){4-6}
    Original stem & Meaning & $w$ & %
    Original stem & Meaning & $w$ \\
    \midrule
    \emph{clandestin}- & illegal immigrant &  -1.85 & %
    \emph{diritt}- & rights & 1.13 \\
    \emph{Cont}-   & PM Giuseppe Conte & -1.30 & %
    \emph{mar}- & sea & 0.89 \\
    \emph{sinistr}-   & the left & -1.15 & %
    \emph{potenz}- & powers & 0.79 \\
    \emph{leg}-   & Lega & -0.90 & %
    \emph{pagin}- & pages & 0.77 \\
    \emph{occup}-   & occupy & -0.89 & %
    \emph{zitt}- & silenced & 0.75 \\
    \emph{sbarc}-   & boat arrivals & -0.86 & %
    \emph{odi}- & hate & 0.73 \\
    \emph{sindac}-   & mayor & -0.84 & %
    \emph{sfrutt}- & exploited & 0.72 \\
    \emph{traffic}-   & trafficking & -0.81 & %
    \emph{problem}- & problem & 0.70 \\
    \emph{ONG}   & NGOs & -0.80 & %
    \emph{turc}- & Turkey & 0.67 \\
    \emph{sol}-   & jus soli\footnotemark & -0.77 & %
    \emph{social}- & social & 0.67 \\
    \bottomrule
  \end{tabular}
\end{table}
\footnotetext{Birthright citizenship in Latin.}

In Table~\ref{tab:ads_example_3}, we report the most predictive features as recognized by our leaning classifier applied on the whole data set of relevant ads.
Among the most predictive features, we find the term \emph{``clandestino/a''} (illegal immigrant); this term has been central in the anti-immigration narrative in Italy~\cite{carlotti2020populists}, since ``where immigrant groups are seen as predominantly illegal, then multi-culturalism poses perceived risks''~\cite{kymlicka2010rise}.
On the other side, words related to humanitarianism are among the most predictive of pro-migration sentiment, such as \emph{``sociale''} (social), and \emph{``diritti''} (rights).
Boat arrivals, despite playing a minor role in Italian immigration, are typically adopted as target by the right~\cite{geddes2020italian} and among the most correlated words with anti-migration sentiments.
Anti-migration messages seem to often mention opponents, associating them with the pro-migration stance: this is the case for ``the left'' and PM Giuseppe Conte.
NGOs who rescue migrants at sea are often the target of anti-immigration attacks, and as such are often mentioned and accused of aiding human traffickers.
On the other side, pro-migrant messages focus on their rescuing activity at sea.
Finally, important signals for pro-migration sentiments are mentions of problem faced by immigrants, such as hate and exploitation.
A more thorough qualitative examination of the content of these ads is left for future work, and instead we focus on their audience reach.

\subsection{Audience demographics}
\label{sec:audience-demographics}

Next, we turn to the audience of these ad campaigns.
Recall that the Facebook Ads Library provides summary demographic statistics on the ``impressions'' received by each ad, as a distribution over 3 genders (male, female, unknown) and 7 age ranges. 
As such, this metric describes the views of each ad, which may be different from users exposed to the ad, as the same user may be exposed multiple times. 
Still, the impressions allow a glimpse into the combination of targeting by the advertisers and the internal Facebook ad display mechanism, resulting in an uneven distribution of each ad among Facebook users.
Using this information, we ask, \emph{what are the difference in the audience of ads from different sources, including different political parties, while taking into consideration the stance these ads express toward the migration issue?}

Out of the 2312 migrant-related ads, anti-migration ads were 677, pro-migration were 1112, while the others 523 were neutral or irrelevant. Considering only ads from major political parties, we have 773 ads, of which 368 anti-migration (47\%), 309 pro-migration (39\%) and 96 neutral or irrelevant (12\%).

Among the 35M impressions received by the migrant-related ads we have collected, those posted by political authors account for 24.5M (69.5\%), followed by NGOs with 2.9M (8.3\%). 
Other account types we were able to identify include universities with 232K impressions, trade unions with 118K, and fact checkers, news agencies, and journalists with 61K impressions. %
Thus, we focus on the ads posted by the political authors and those from NGOs, as they account for the majority of impressions, and, as we see shortly, provide opposing stances.

\begin{figure*}[t]
    \centering
    \includegraphics[width=\textwidth]{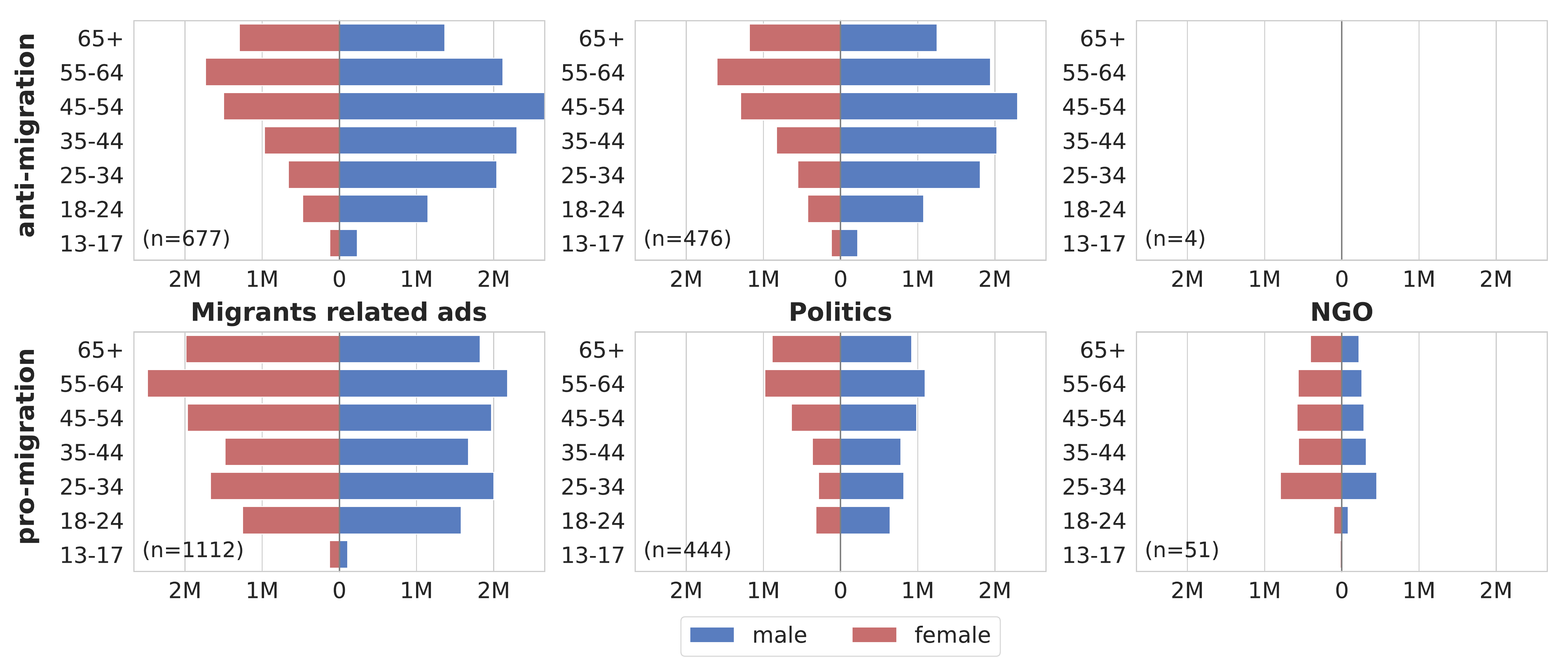}
    \caption{Distribution of impressions over age and gender of (left) all migrant-related ads, (center) all ads from political authors, and (right) NGO authors, grouped by pro-migration and anti-migration classification. Note, all x axes have the same range. In each plot, we report the total number of ads represented as $n$.}
    \Description{See text for a detailed description.}
    \label{fig:gender_age_pyramid}
    \vspace{-\baselineskip}
\end{figure*}

Figure~\ref{fig:gender_age_pyramid} shows the demographic distributions over age and gender (unknown gender dropped due to having almost no activity) for all migration-related ads (left), political ads (center), and those by NGOs (right).
We show the total impression by ads classified as pro-migration on the bottom, and by anti-migration ads on top. 
Considering all ads, the pro-migration ones are somewhat more symmetrical in gender distribution than the anti-migration. 
Whereas anti-migration advertising is favoring older users and males, the pro-migration advertising is viewed mostly by females. In particular, the odds ratio for this gender distribution difference is 1.69 (considering the whole dataset), meaning that an anti-migration ad is 69\% more likely to be seen by a male user than a female and a pro-migration one.
We note that the female part of the audience of anti-migration ads is even more skewed towards older age than males.

These trends are largely due to the ads coming from the political advertisers (center), who account for almost all of the anti-migration advertising captured in the dataset. 
Conversely, NGOs post only pro-migration ads.%
\footnote{The classifier did identify 4 NGO ads as anti-migration, but upon manual inspection we found all to be mis-classified.}
These ads, unlike those posted by political authors, are seen mostly by younger women. 
Thus, we find NGOs to assume a clear stance on the migration debate.
Instead, different ads from political authors can represent different opinions.

\begin{figure*}[t]
    \centering
    \includegraphics[width=0.9\textwidth]{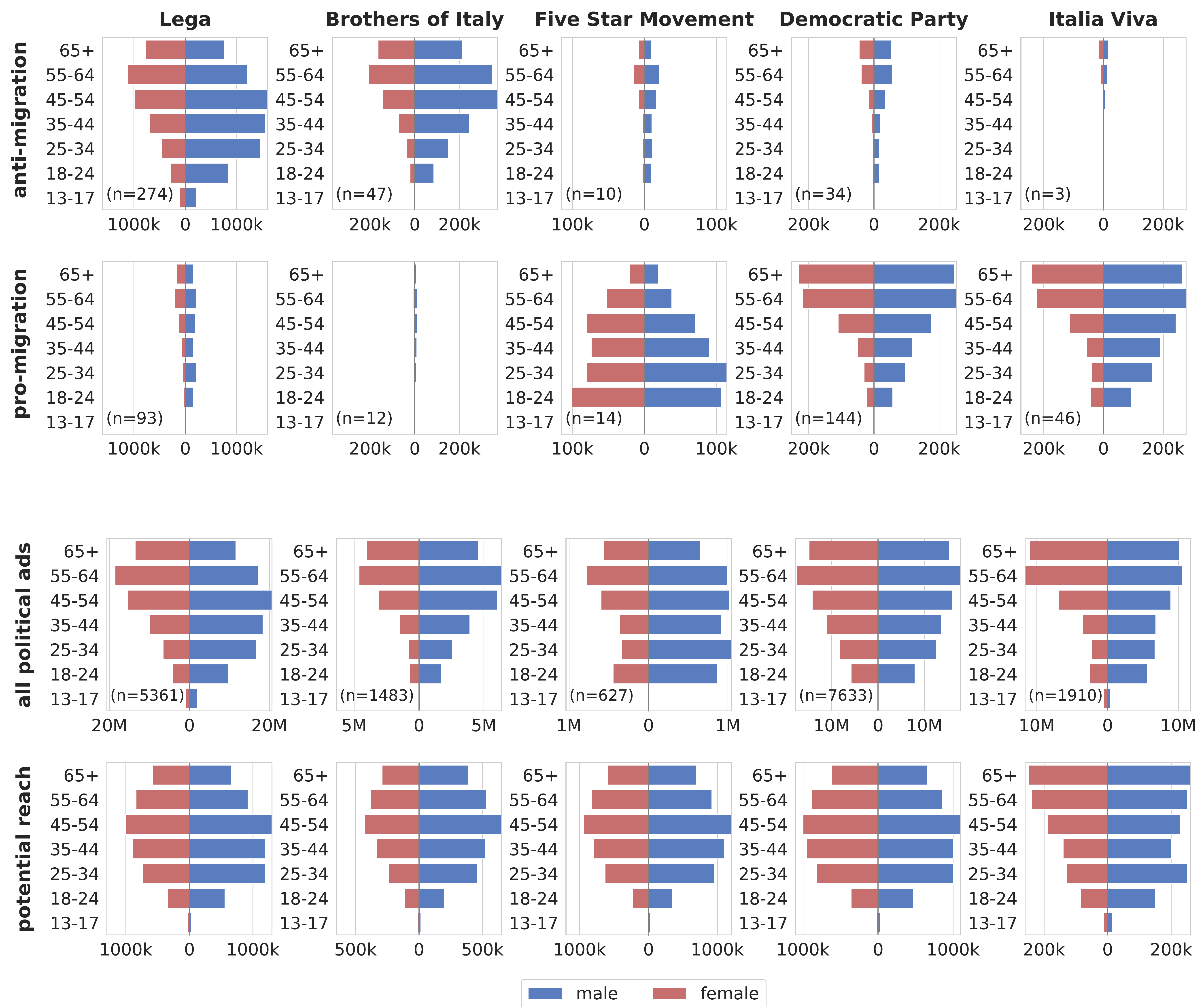}
\caption{Distribution of impressions over age and gender of ads posted by five political parties, for pro-migration ads, anti-migration ads, all political ads, and the potential reach on the platform. Note, each party has distinct x axis range; the first two lines (pro- and anti-migration ads) share the same axis. In each plot, we report the total number of ads represented as $n$.}
    \label{fig:gender_age_pyramid_parties}
    \Description{See text for a detailed description.}
\end{figure*}

These two sides in the political domain can be attributed to the different party affiliations, as Figure \ref{fig:gender_age_pyramid_parties} illustrates. 
Here, we show again the audiences for anti- and pro-migration ads, as well as two ``baselines'': impression distribution for all political ads by the party, and finally party's potential reach on Facebook (see next paragraph).
Out of the five major political parties, we find Lega and Brothers of Italy to mostly post ads with an anti-migration stance, whereas Five Star Movement, Democratic Party, and Italia Viva with pro-migration one. 
Despite the stance, most parties achieve more male viewership than female. Considering this subset of ads by major parties, the odds ratio for this gender distribution difference is 1.17, meaning that a migrant-related ad ad is 17\% more likely to be seen by a male user than a female. Also, we found that anti-migration parties focus on more male audience specifically for their migration ads with respect to ads with other topics, with an odds ratio of 1.24: an ad from these parties is 24\% more likely to be seen by a male user when the ad speaks about migration, than if it does not.
Five Star Movement has the most gender-equal representation, and is the only one reaching a younger audience.
This is consistent with the history of M5S of focusing on the younger part of the population~\cite{natale2014birth}.
However, it is interesting to note that the very few ads from M5S that have been classified as anti-migration show a completely different demographic, significantly skewed towards older people.
Upon manual inspection of a few of those ads, they appear to be correctly classified by our approach, as they (implicitly) convey an anti-migration sentiment.
This seems consistent with the idea that inside the Five Star Movement, different currents coexist~\cite{natale2014birth}, including former left-wing activists and former right-wing voters, which leads to wavering messages on the topic of migration~\cite{pasini20196}.
We show here that these messages are, in the end, reaching different parts of the electorate thanks to the Facebook advertising system.
The Democratic Party seems also to be engaging in different messages, even if their demographic distribution appear to be quite similar in the two cases.
However, overall, each party is clearly associated with one of the two sides.
In general, our results show that polarization about migration tend to fall along party lines, at least in the message conveyed by each party, which is consistent with previous reports of polarized attitudes in the media on the topic~\cite{van2017will}, and with the polarization of the issue in Italy during the considered time frame~\cite{geddes2020italian}.

\paragraph{Comparison Facebook audience interested in a party.}

We also consider the potential Facebook audience of each party (last row of Figure~\ref{fig:gender_age_pyramid_parties}),  retrieved via Facebook Ads Manager,\footnote{\url{https://www.facebook.com/ads/manager}} using Monthly Active Users who are in Italy and are interested in the particular party (collected on August 6, 2020).
Intuitively, if the parties were to simply target people who are interested in their party, this would be the distribution of people they would reach.
We observe that the potential reach of the major parties is similar to one another, while their effective audiences differ.
This observation suggests that parties are, intentionally or not, more effective in reaching specific segments of the population.

\paragraph{Comparison with surveys.}

We also compare these audience characteristics to those of their voter base. %
For each party, we compare the distribution of impressions of ads over age groups (genders) to the voter demographics as estimated via a large IPSOS survey \cite{ipsos2019elezioni}.
However, in order to perform this comparison, we need to face two issues.
First, age buckets are different between the two data sources: in the \textsc{IPSOS} surveys we have the four buckets 18-34, 35-49, 50-64, and 65+, while from Facebook Ads Library we have the seven age buckets reported in Figure~\ref{fig:gender_age_pyramid_parties}.
Therefore, we have to choose the three buckets for which the borders of each bucket are the same in the two sources, that is 18-34, 35-64 and 65+.
We also note that the surveys do not show the interaction between gender and age; also, surveys did not report Italia Viva, as it was not yet split from PD at the time of elections.

Our second challenge is that we wish to compare Facebook impressions to statistics about the general population; thus, we introduce a conversion factor which we apply to the impression data to make it comparable to the voter demographics.
Let the impressions achieved by all ads of a party $p$ in a demographic group $g$ be $A_{g, p}$ and $P_g$ be the proportion of the demographic group $g$ in Italian population.
We also consider the potential Facebook audience $F_{g, p}$ of each party $p$, as retrieved via Facebook Ads Manager and operationalized by using Monthly Active Users who are in Italy and are interested in the particular party $p$ (collected on August 6, 2020). 
We compute the conversion factor $\frac{P_g}{F_{g, p}}$ which can be used to compare the Facebook impressions to voter demographics.
This way, the converted estimate $A'_{g, p} = A_{g, p} \cdot P_g / F_{g, p}$ can be thought of as a proportion of people reached by the ads, normalized ``as if'' Facebook users were distributed as the general population.
This computation is then performed for each party $p$, thus obtaining a distribution over demographic groups for that party, which then we compare to the demographic distribution of its voters from surveys.

\begin{figure*}[t]
    \centering
    \includegraphics[width=0.9\textwidth]{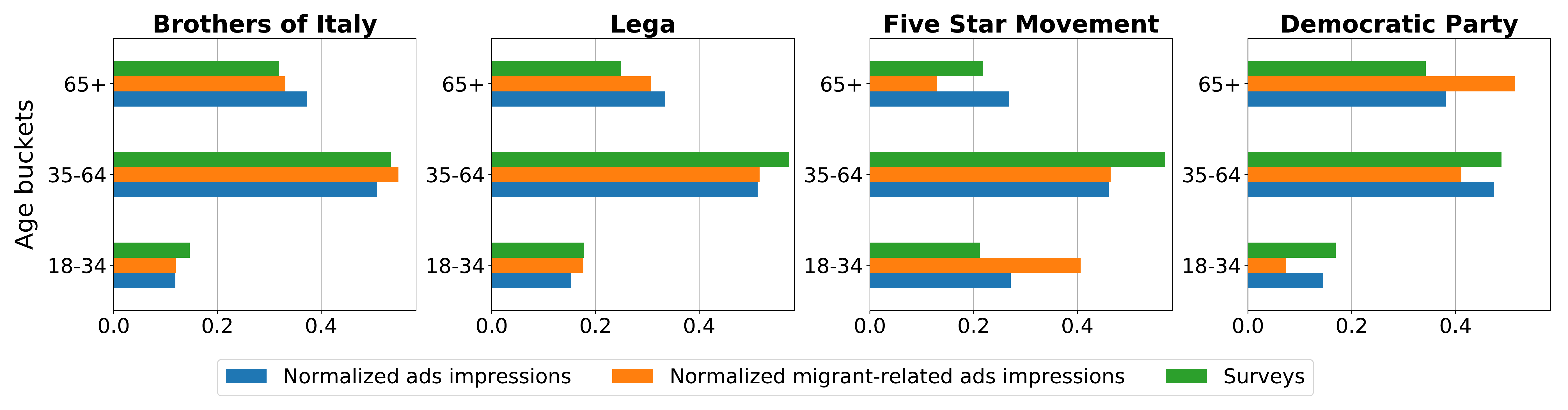}
    \caption{Comparison between age groups in each party supporters, as reported by surveys~\cite{formigoni2018elezioni}, and age groups in the viewership of each party's ads -- all ads (blue, bottom bar) and migration-related ads (orange, middle bar), normalized to remove the effect of Facebook users demographic. Italia Viva is not included as it did not participate in European elections.}
    \Description{The demographic distribution of audience approximately shows the same asymmetries reported by surveys on the general population. For some parties -- Lega and Brothers of Italy -- the two distribution resemble each other much more significantly, while for the others -- M5S and PD -- they differ.}
    \label{fig:imps_vs_survey}
\end{figure*}

We report results about this comparison in Figure~\ref{fig:imps_vs_survey}.
We notice that, in general, the demographic distribution of audience approximately shows the same asymmetries reported by surveys on the general population: for instance, in the 65+ age group, PD is the strongest party while M5S is the weakest; Lega attracts more 18-34 voters than Brothers of Italy.
However, we observe that for some parties -- Lega and Brothers of Italy -- the two distribution resemble each other much more significantly, while for the others -- M5S and PD -- they differ.
Since the division reflects the one we observed in Figure~\ref{fig:gender_age_pyramid_parties} between pro- and anti-migration ads, the two phenomena might be related.
For instance, a possible explanation is that the anti-migration parties are using Facebook ads mostly to reach the demographic more sensible to this message, while for parties that are less invested on the issue the audience for migration-related ads is less representative of their voter base.

\subsection{Audience targeting}
\label{sec:ads-targeting}

\begin{figure}
\centering
\begin{minipage}{.45\textwidth}
  \centering
  \vspace{0.5cm}
  \includegraphics[width=\linewidth]{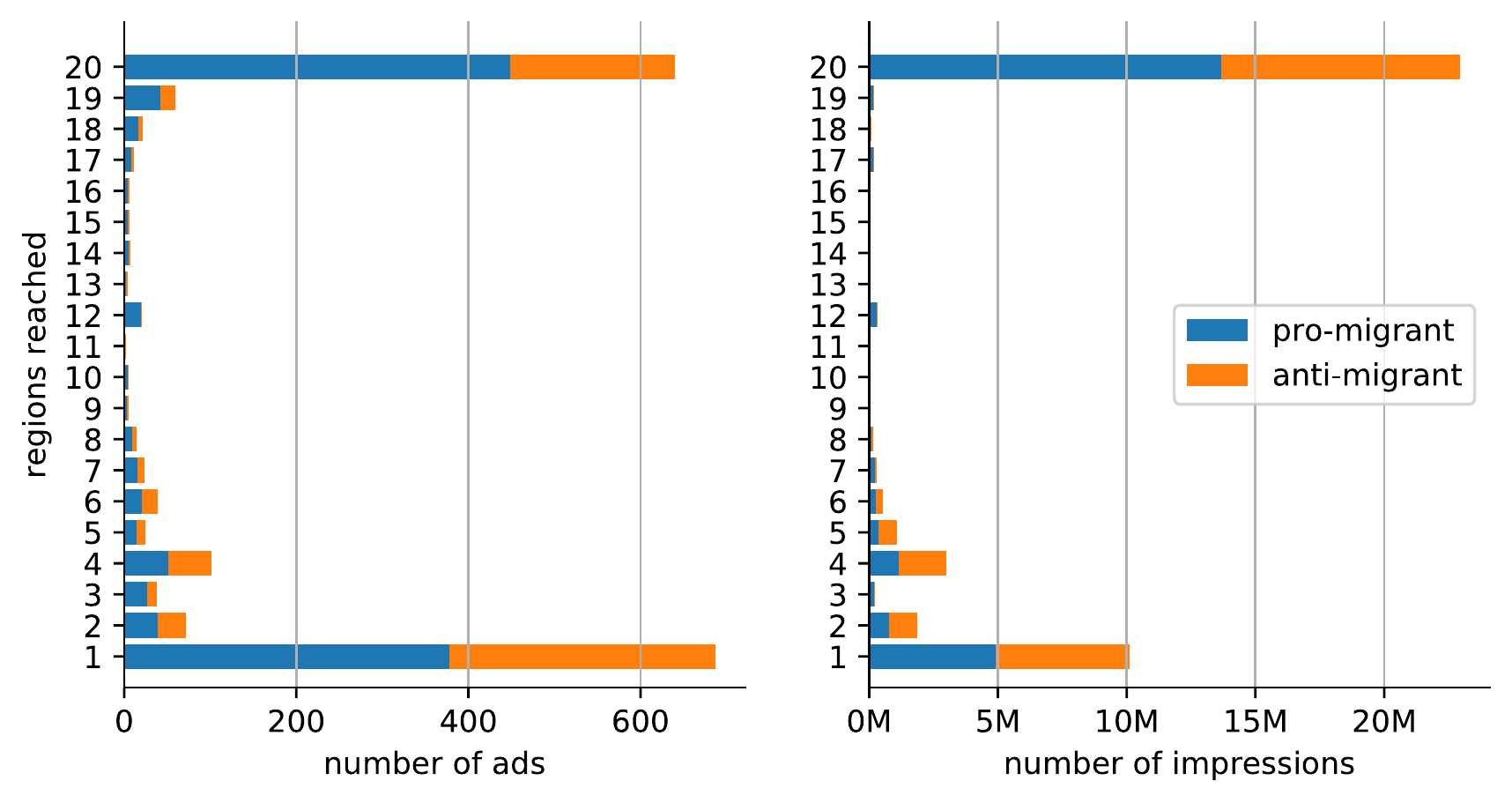}
  \captionof{figure}{Geographical Targeting.
  Left: barplot with number of ads that reach a certain number of regions in Italy. Roughly half the ads are targeted to a few regions, while the others reach most of the regions and don't use geographical targeting.
  Right: the number of impressions produced by ads that reach a certain number of regions.
  }
  \label{fig:regions_reached}
\end{minipage}
\begin{minipage}{.45\textwidth}
  \centering
  \vspace{0.5cm}
  \includegraphics[width=\linewidth]{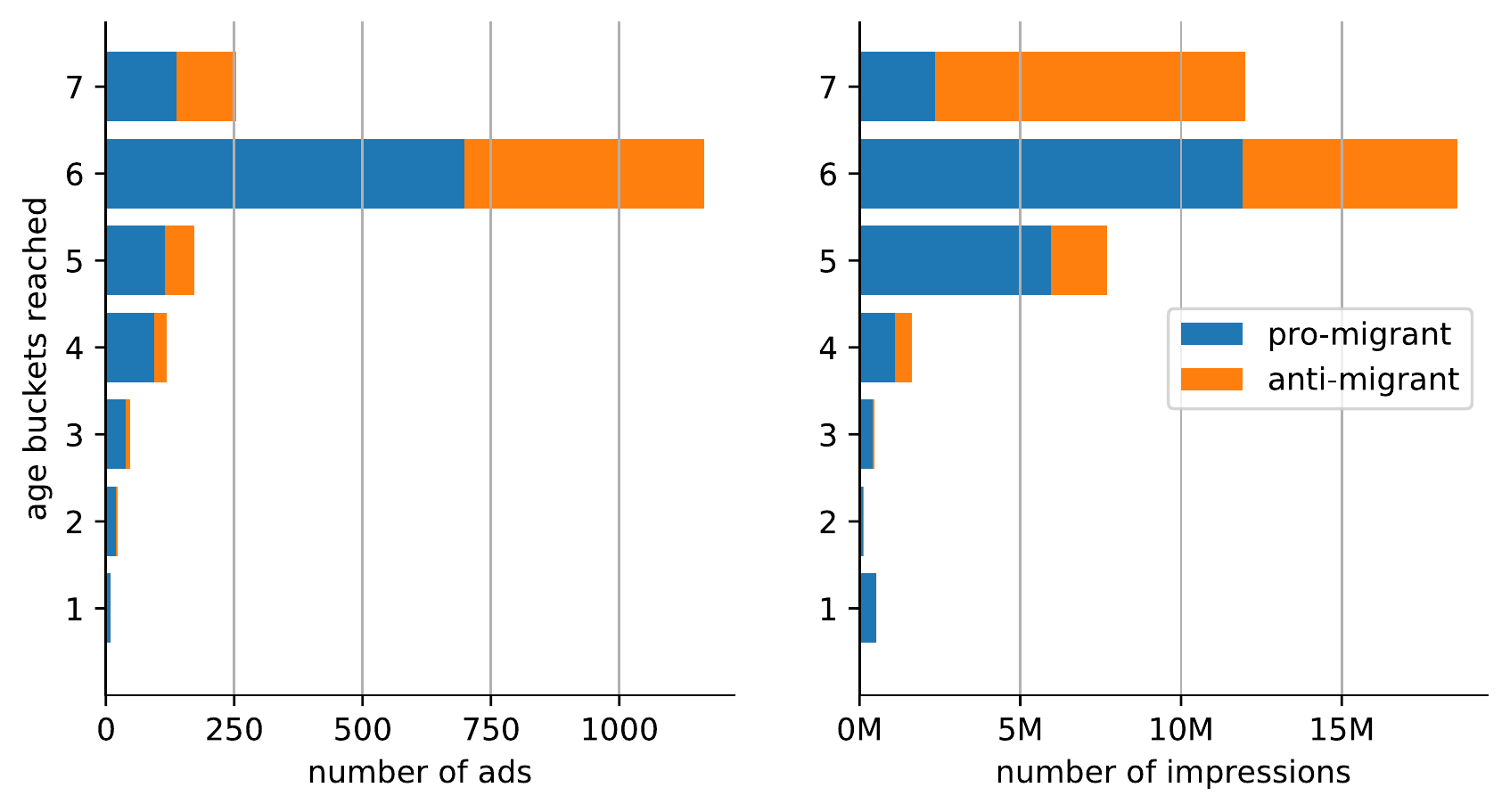}
  \captionof{figure}{Age Targeting.
  Left: barplot with number of ads that reach a certain number of age buckets. Just a few ads target a single age bucket, while most ads exclude just one age bucket (most commonly the 13-17 bucket).
  Right: the number of impressions that come from ads that target a certain number of age buckets.
  }
  \label{fig:ages_reached}
  \Description{Roughly half of the ads use geographical targeting. Most of the ads do not reach all 7 age buckets, mostly because they avoid to target the 13-17 age bucket.}
\end{minipage}%
\end{figure}

Next, we perform a quantitative analysis over the ads to check how many intentionally use geographical, age, or gender targeting such that some demographic group is totally excluded from the audience.
Specifically, we check for each ad how many of the 20 Italian regions are reached by that ad (geographical targeting), how many of the 7 age buckets are reached (age targeting), and if the ad is targeted only to males or to females (gender targeting).
Figure~\ref{fig:regions_reached}
shows that roughly half of the ads use geographical targeting (i.e., audience is not spread in all 20 regions). The ads with higher number of impressions do not use geographical targeting, but at least a third of the impressions comes from ads with such geographical targeting.
Such targeting could be explained by local elections.
Figure~\ref{fig:ages_reached} shows that most of the ads do not reach all 7 age buckets, mostly because they avoid to target the 13-17 age bucket. Most of them reach all other 6 buckets, with no age targeting. However, if we sum the impressions, the ads with highest number of impressions also reach the 13-17 age bucket, and a good part of the impressions come from ads with at least some degree of age targeting (i.e., exclude at least one of the other age buckets).
Finally, we find just a few  ads (18) that are targeted exclusively towards male or female users; they are responsible for less than 0.2\% of the total number of impressions.

In total, if we consider ads that have impressions on all regions, all age buckets (except the 13-17 age bucket) and all genders as ads that do not use demographic targeting, we find that 38\% of the impressions are produced by these ads, while 62\% of the impressions are produced by ads that use some sort of demographic targeting.

\subsection{Temporal relationship with traditional news media}
\label{sec:news}

\begin{figure*}[t]
    \centering
    \includegraphics[width=0.8\textwidth]{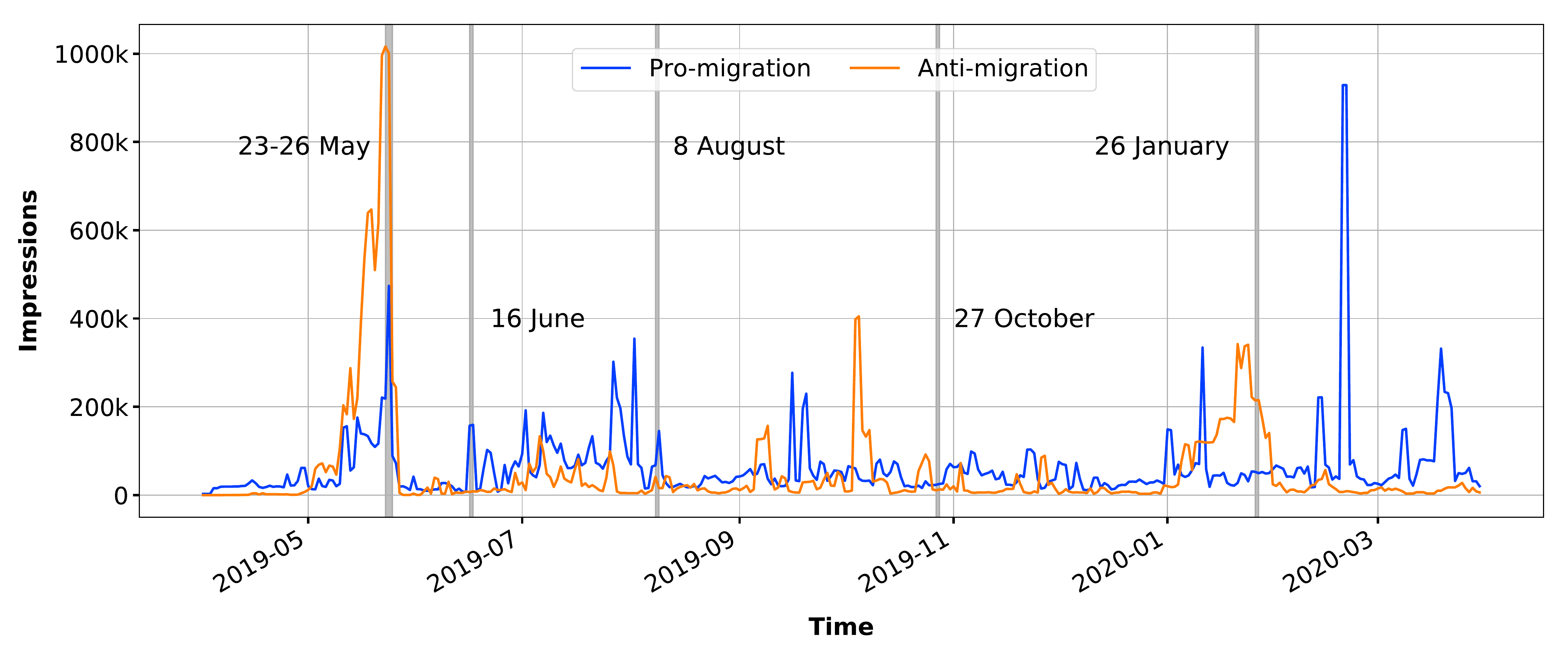}
    \caption{Impressions for pro- and anti-migration ads per day. We also highlight important political events with vertical bars: 23-26 May - European and regional (Piedmont) elections; 16 June - Regional elections in Sardinia; 8 Aug - Lega leaves the government; 27 Oct - Regional election in Umbria; 26 Jan - Regional elections in Calabria and Emilia-Romagna.}
    \label{fig:pro-anti-over-time}
    \Description{Impressions for pro- and anti-migration ads per day, highlighting large spikes often accompanied by political events.}
\end{figure*}

After inspecting the demographics of \emph{who} is the audience of these Facebook ads, we now move our attention to \emph{when} the audience is exposed to them.
When we consider the temporal evolution of these campaigns (Figure \ref{fig:pro-anti-over-time}) and the attention they receive, we find large spikes.
As the figure indicates, many of such spikes happen around major political events such as elections, yet others seem not to have an associated prominent event.
As found by~\citet{thorson2018attention}, important political events increase the effectiveness of political advertising in capturing the attention of the public.
For instance, the first peak happens during 2019 European elections.
We see both pro- and anti-migration messages increase in the weeks before these elections.
Right after the elections, the attention drops, since all parties focus their efforts on the campaign before it.
For most other peaks we find instead that they are caused by only pro- or only anti-migration messages -- upon further inspection we find that they are caused by one single party, often trying to emphasize specific news or events.
Interestingly, we also find that the large peak in January 2020 happens right before the regional elections in Emilia-Romagna -- and, in fact, presents the same sawtooth-like shape -- but is instead formed only by anti-migration messages.
This happened since elections in Emilia-Romagna were important on a national level for Italian politics~\cite{emiliaromagna}: the leader of Lega, Salvini, tried to campaign heavily on anti-immigration messages to challenge the current government~\cite{amaro2020salvinis}, in a region historically dominated by left-wing politics.
Finally, the last peak is a single ad by Renzi (leader of Italy Alive) that attracted plenty of attention by capitalizing on his intervention in the senate, linked as a video in the ad.
The ad included in its brief text a sentence about migration, that translates as ``Italy shall not give money to whom does not welcome migrants, such as Hungary''.

\paragraph{Granger causality with news.}
Since, as we observe large spikes, some of which are accompanied by a political event, it is natural to ask whether ads impressions are more likely to follow major news news (``riding the wave''), or whether there are more instances of them following their own internal timing (``owning the issue'') and potentially impacting the news.
To check this, we compute two time series:

\begin{enumerate}
    \item For each article $i$, we compute $\frac{T^m_i}{T_i}$ as the fraction of themes of article $i$ that are related to migration. In order to define our time series $\textsc{News}(t)$ for each day, we sum these fractions for all the articles $D_t$ that were published on that day: \begin{equation}
        \textsc{News}(t) = \sum_{i \in D_t} \frac{T^m_i}{T_i}
    \end{equation}
    \item For ads, we compute the total number of impressions of migration-related ads for a given day, $\textsc{AdsImpressions}(t)$, by uniformly distributing the number of impressions across the duration of each campaign.
    We also consider only pro- and anti-migration ads.
\end{enumerate}

With these two time series defined, we compute the following two Granger-causality tests: one to check if $\textsc{News}(t)$ Granger-causes $\textsc{AdsImpressions}(t+\delta)$, and one in the opposite direction.
The hypothesis assumed by the first test is that migration-related ads receive more attention after news discuss migration-related issues, consistently with the ``riding-the-wave'' theory~\cite{ansolabehere1994riding}.
The second test instead hypothesizes that news follow the migration topic after related social media political ads receive more attention, implying that political ads are able to ``set the agenda'' for mainstream media~\cite{roberts1994agenda}.
We exclude from the analysis the period after December 2019, since the unprecedented situation related to the beginning of the COVID-19 outbreak in Italy could affect results.

\begin{figure*}[t]
    \centering
    \includegraphics[width=0.7\textwidth]{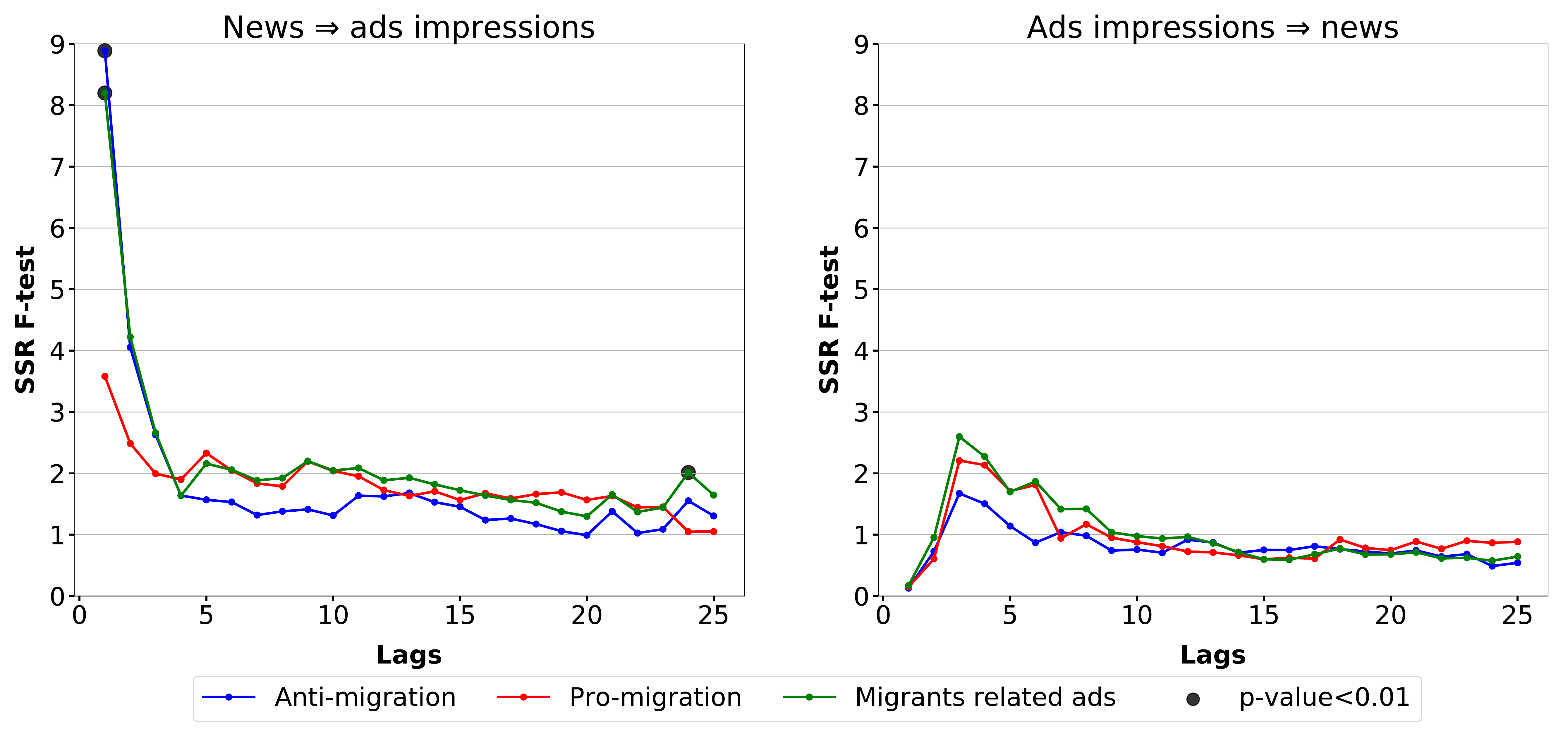}
    \caption{Granger-causality tests comparing migration-related news and ads impressions, for the set of all migration-related ads, and for pro- and anti-migration ones only, up to December 2019. %
    On the X axis, we report the number of lags in days we consider for the delay between the two time series; on the Y axis, we show the sum-of-square F-test statistics for that lag. We highlight those with P-value < 0.01.}
    \label{fig:prob_age_gender_baseline}
    \Description{Granger-causality tests comparing migration-related news and ads impressions. We see a significant F-test for the hypothesis of News Granger-causes Ads at offset of 1 day.
    We find no significant Granger-causality from ads impressions to news.}
\end{figure*}

We report results for these two tests in Figure~\ref{fig:prob_age_gender_baseline}, for the case where we consider all migration-related ads, and for the pro- and anti-migration ads only.
We see a significant F-test for the hypothesis of $\textsc{News}(t)$ Granger-causes $\textsc{AdsImpressions}(t+\delta)$ at offset $\delta=1$ day.
This result is significant also by considering only anti-migration ads, but not when considering pro-migration ones.
Conversely, we find no significant Granger-causality from ads impressions to news.

This finding therefore supports the riding-the-wave hypothesis: when news are focusing more on migration, then Facebook ads -- and especially anti-immigration ones -- are ``jumping on the bandwagon'' and attracting more attention.
This $\delta=1$ day lag could be an artifact due to news following a different publication cycle than Facebook ads; however, in practice Facebook allows for ads to be published at any moment, making it more likely that news indeed sets the agenda for the ads, especially for some political parties, which are responsible for the large majority of anti-immigration ads.

\section{Discussion}

Internet advertising is a powerful engine that finances the existence of online content, and allows a personal connection between advertisers and potential viewers.
Human interaction with advertising has been an important part of HCI research, from detecting advertising campaigns~\cite{swart2020ad}, to revealing privacy concerns~\cite{qalbi2020aranyani}, to understanding ad framing~\cite{pak2017news}.
Our analysis describes how politicians and organizations use digital platforms such as Facebook to promote their own agenda.
As such, this work is at the intersection of technology and social science.
The HCI community has devoted a great deal of effort into designing algorithms that use behavioral targeting to make advertisements more effective, as well as analyzing how companies use them and how these ads are received by users.
For instance, \citet{eslami2018communicating} found that ``users preferred interpretable, non-creepy explanations about why an ad is presented'' on Facebook.
Since we find that political advertisers disproportionately reach some demographic groups, this information should complement ads explanations -- and possibly be part of public policies on ads design.
This finding is particularly important since \citet{johnson2018analyzing} showed that ``consumers cannot always distinguish paid from unpaid content''.
Suggestions have been made to turn this vast machinery to the benefit of the viewers, such as designing personal goal-oriented ads~\cite{ramsay2019yourad}, allowing proceeds to go to environmental causes~\cite{qalbi2020aranyani}, or even counteracting the influence of seen ads~\cite{pinder2017anti}.
For example, \citet{ali2019discrimination} showed how gender and racial bias could be a product of an apparently neutral design of ads.
In our work we show how such choices are employed in practice in political campaigns that specifically talk about migrants, a particularly vulnerable group.

In general, the possibilities offered by this data are endless, for example to explore the design and implementation of digital large-scale campaigns on controversial topics, on one hand, and the exposure and fruition by specific population groups on the other, mediated by the technological interface of a digital social-media/advertisements/communication platform.
Until recently, there were limited resources to track whole advertising campaigns, as this information was not provided by the platforms.
Facebook Ads Library is a new tool that lets us monitor advertising around social issues on the largest social network on the Internet, and this paper proposes a pipeline to process this data and showcases the insights it can provide.
Nevertheless, this data source is not perfect.
For instance, while the final reached demographic of each ad campaign is known, the reason why the ad was shown to this specific group is undetermined:
it could be either because of explicit targeting by the author, or because of bias in the Facebook optimization algorithm.
To answer this type of questions, the HCI community needs to work with the data providers to help improve their transparency efforts.

In our dataset, we found over two-thousand advertising campaigns that were targeting Italian users of Facebook on the topic of migration and refugees--ads that gathered millions of impressions. 
Upon classifying these ads as pro- and anti-immigrant, we found signs of strict polarization:
the division between anti- and pro-migration ads, as detected by our classifier, clearly splits different groups of advertisers.
NGOs, often attacked by anti-immigration politicians because of their rescuing activities in the Mediterranean, use Facebook to circulate pro-migration messages.
For instance, we found NGOs promoting voluntary activities for immigrant refugees, or meetings with them to understand their backstories. %
Different political parties mostly picked one side: Lega and Brothers of Italy on anti-immigration side and Five Star Movement, Democratic Party, and Italia Viva on the other.
Out of all the migration ads by these five parties, we find 47.6\% of them had anti-migration slant, and only 39.9\% pro-migration, but the anti-migration ones account for 65.2\% of the impressions, whereas pro-migration only 27.0\%, which contributes to the earlier observations of the social media tendency to favor negative views and anger-inducing messages~\cite{hasell2016partisan,lee2018does,tucker2018social,hannan2018trolling}. 
Note, however, that these stances are quite close to those revealed by the Italian Eurobarometer survey in 2019, where to the question of ``what is your opinion on immigrants to Italy from non-EU countries?'', 41\% responded positive, and 51\% negative (the rest neutral)~\cite{eurobarometro2019italia}. 
The content of anti-migration ads spans many topics that may be interesting to different audiences: accusation of NGOs working with human traffickers, laws concerning immigration, and security issues, as well as more general discussion about defending Italian culture and the role of the government and its priorities. 
Although outside of the scope of this study, a qualitative examination of the issue framing would provide a fine-grained view of the leverage used during the political discourse via advertising.

When looking at the demographic characteristics of pro- and anti-migration ads, we find that NGOs reached mostly women rather than men with their pro-migration message, while the opposite is true for all political parties.
However, the audience of anti-migration ads by political parties is more skewed toward males, both with respect to pro-migration ads and to the general audience of their ads on other topics.
We also find that targeting that explicitly excludes one gender is rare in our data set (0.2\% of the impressions), meaning that such a skewed distribution is likely to be the by-product of a marketing optimization by the platform; for instance, males could be targeted with anti-migration ads since they are more likely to be interested in that message.
Furthermore, we find that the anti-migration ads from political parties are being shown to an audience that resemble their voters in age distribution.
This is not the case for pro-migration ads from the other parties: their audience resembles their voters only when considering all their ads, but not while singling out the migration-related ones.
This finding is coherent with the idea that the anti-migration message is central to some parties, and therefore users interested in that message resemble the whole, while pro-migration parties have other main focuses.

In some instances, the audience reached by migration-related ads differed from the usual audience of the party, such as in the striking example of the Five Star Movement, who reaches younger users disproportionately more with their pro-migration ads than with other ads.
The opposite is true for the Democratic Party, who reaches older users with pro-migration ads than with other ads. 
Some groups of users could be targeted more by messages on a specific topic if that message is likely to resonate with their views.
Interestingly, the few impressions that the Five Star Movement obtains from anti-migration ads -- since the overwhelming majority of their impressions are from their pro-migration ads -- shows a totally different age distribution, much more skewed towards older users.
This could suggest that some limited campaigns (e.g., from local politicians) might display different messages to different targets, when compared to the global audience of the same party on the same topic.

The above findings suggest some key points in terms of design practices, that could help increase transparency and accountability.
Considering the importance of providing interpretable explanations to users~\cite{eslami2018communicating}, users should be able to access clear explanations about \emph{why} some demographic groups are being reached by an ad.
In our findings, we cannot distinguish between different causal explanations for the demographic asymmetries we find among ads: they could be obtained on purpose by the advertiser, or they could be a by-product of the auditing algorithm operated by the platform.
If the Facebook Ads Library were to surface more information on the targeting configuration of socio-political campaigns, it would greatly elucidate this matter.

First, additional targeting settings (that are currently not revealed by the Library) would make clear whether the advertisers are targeting minorities of race or sexual orientation, or those having particular beliefs or tendencies (such as nationalism, anti- or pro-European Union stance, etc.).
For instance, on the Facebook's Advertising platform an advertiser can constrain the audience of an ad to those users who have politically-relevant interests in ``European Union'', ``National Security'', or even specifically ``Refugees of the Syrian Civil War'', or others such as ``LGBT community'', ``Human Rights Watch'', or ``Second Language''. 
These interests can be used both to target individuals, but also to exclude users from the audience. 
Thus far, few rules have been established to curb targeting by the political advertisers, although some efforts are ongoing, such as those by the US Federal Communications Commission, which requires the disclosure of the sponsoring entity,\footnote{\url{https://www.fcc.gov/media/policy/statutes-and-rules-candidate-appearances-advertising}} and by the Italian Data Protection Authority, which constrains the use of personal data by platforms.\footnote{\url{https://www.garanteprivacy.it/web/guest/home/docweb/-/docweb-display/docweb/9105201}}
Second, social media platforms have an enormous power to shape the advertising campaigns by enforcing their policies that may concern hate speech, misinformation, or involve vulnerable populations.
In fact, as private companies, they are free to enforce any internal policy they choose~\cite{nott2020political}.
Interviews with insiders and employees of these companies~\cite{kreiss2019arbiters,keller2020facts} show that these policies are determined via ``internal debates, appeals by practitioners, and outside pressure'' and are applied in opaque ways.
Unfortunately, the Library does not reveal the advertisements that have been rejected by Facebook.

If we do assume that most advertisement is shown to the people who are already interested in the parties and who are sympathetic to their message (which is likely valid for many such impressions), we can view these ads as an extension of the opinion ``echo chamber''~\cite{garimella2018political,del2016echo} wherein an individual's viewpoint is reinforced by the interaction with similar-minded others. 
In that sense, the reinterpretation and re-framing of current events via internet advertising allows the politicians to produce time-relevant content that supports and reinforces their message.
Indeed, we find that the advertising impressions closely follow the mainstream news volume around the topic of migration, ``riding the wave'' of attention to the ongoing news.
Yet the stance of each party's message remains constant over time, with each news item being re-framed by the advertisers to suit their stance.
Briefly, we attempted to compare the themes in the Italian mainstream news and ads around the same time, but found that the differences in vocabularies (and their sizes) introduce too much noise to provide a reliable signal. 
However, the framing of ongoing events by social media advertisers is an exciting direction of future research, especially since much of advertising is increasingly shown as ``organic'' content, which is difficult to distinguish from legitimate social media posts.

Although Facebook Ads Library gives an unprecedented view of the political campaigns on the largest social media platform, our insights are still limited in several ways. 
First, the selection of keywords defines the scope of our analysis, and there are surely ads relevant to migration that slipped through our filter, potentially impacting non-mainstream messages.
Second, as we discuss above, only very general information is available on the audience reached.
In particular, ``impressions'' does not equal the number of people reached, as one user may see the ad on a number of occasions. 
Neither can we compute the overlap in viewership of different campaigns, which is surely substantial within the those interested in a particular party.
Third, despite Facebook being a mammoth among increasingly popular social media, it is by far not the only communications outlet available to the politicians and other actors. 
Traditional local news media, as well as new online newspapers such as \texttt{Fanpage.it} still account for the bulk of news consumption in Italy~\cite{cornia2019digitalnews}.
Finally, the interaction with such ads likely excludes many of those with disabilities preventing the full use of the platform, such as those with visual impairments~\cite{glaser2019blind}, constraining further the population that is reachable.

Despite many limitations, Facebook Ads Library still provides exciting possible future directions of research. 
A regular, daily crawl of the data would allow for a finer-grained view of the viewership statistics of the ongoing campaigns, and the conditions at which the campaigns are terminated.
Connection to additional communication channels of major advertisers would allow for a broader view of the campaign, and especially would support comparative analysis of messaging in different media.
Finally, a more qualitative analysis of the ad content would allow for a more nuanced breakdown of psychological tactics, re-framing of news events, and calls to action that are employed by the advertisers.
The success of these can then be measured by cost per impression.
Moreover, studies such as this one are necessary to examine the accountability efforts of major social media platforms in response to increasing criticism~\cite{sunstein2018republic} (such as Google's Transparency Report~\cite{google2020transparency} and Twitter Ads Transparency Center\footnote{\url{https://ads.twitter.com/transparency}}).
The completeness and usefulness of this data must be continuously scrutinized in order to support an informed dialogue between the public and the private spheres of the communication ecosystem.

\begin{acks}
The authors acknowledge support from the Lagrange Project
of the Institute for Scientific Interchange Foundation (ISI
Foundation) funded by Fondazione Cassa di Risparmio di
Torino (Fondazione CRT).

\end{acks}

\balance
\bibliographystyle{ACM-Reference-Format}
\bibliography{migrationads}

\end{document}